\documentclass[aps,nofootinbib,twocolumn,showpacs,superscriptaddress,groupedaddress,floatfix,prd,amsmath,amssymb,longbibliography]{revtex4-2}

\usepackage[english]{babel}

\usepackage[T1]{fontenc}
\usepackage[utf8]{inputenc}
\usepackage{amsmath,amssymb,amsfonts}
\usepackage{bm}
\usepackage{physics}
\usepackage{graphicx}
\usepackage{hyperref}
\usepackage{xcolor}
\usepackage[normalem]{ulem}

\newcommand{\ii}{\mathrm{i}}
\newcommand{\e}{\mathrm{e}}

\begin{document}

\title{Bound-State Spectra of a Lifshitz-Type Dirac Equation in (2+1) Dimensions}


\author{Lucas K. R. Queiroz}
\email{lucas.queiroz@icen.ufpa.br}
\affiliation{Faculdade de F\'isica, Universidade Federal do Par\'a, Avenida Augusto Corr\^ea 01, 66075-110, Bel\'em, Par\'a, Brazil}

\author{Van S\'ergio Alves}
\email{vansergi@ufpa.br}
\affiliation{Faculdade de F\'isica, Universidade Federal do Par\'a, Avenida Augusto Corr\^ea 01, 66075-110, Bel\'em, Par\'a, Brazil}

\author{Nilberto Bezerra}
\email{jose.bezerra@icen.ufpa.br}
\affiliation{Faculdade de F\'isica, Universidade Federal do Par\'a, Avenida Augusto Corr\^ea 01, 66075-110, Bel\'em, Par\'a, Brazil}

\author{Luis F. F. Aguilar}
\email{luis.fernandez@ufrontera.cl}
\affiliation{Departamento de Ciencias Físicas, Facultad de Ingeniería y Ciencias, Universidad de La Frontera, Avenida Francisco Salazar 01145, Casilla 54-D, Temuco Chile}

\author{Francisco Peña}
\email{Deceased}
\affiliation{Departamento de Ciencias Físicas, Facultad de Ingeniería y Ciencias, Universidad de La Frontera, Avenida Francisco Salazar 01145, Casilla 54-D, Temuco Chile}

\date{\today}

\begin{abstract}

We investigate a Dirac-type equation in $(2+1)$ dimensions modified by Lifshitz spatial derivatives with dynamical exponent $z=2$, focusing on the spectral properties of bound-states under radial confinement. Analytical solutions are obtained for constant backgrounds, hard-wall confinement, and harmonic potentials, while logarithmic confinement is treated numerically via the Numerov method and complemented by a semiclassical WKB analysis. The resulting spectra exhibit characteristic scaling laws governed by the Lifshitz parameter $b$, including $E - M \propto b/R_0^2$ for hard-wall confinement, $E - M \propto \sqrt{2b}\,\omega$ for harmonic trapping, and $E - M \sim \alpha \ln\sqrt{b}$ in the semiclassical regime of logarithmic confinement, where $M$ is a mass scale. These results provide a consistent characterization of how higher-order spatial derivatives modify the energy spectra in two-dimensional Dirac systems and may be relevant for effective descriptions of materials with quadratic low-energy dispersion, such as bilayer graphene and related anisotropic 2D systems.

\end{abstract}

\maketitle

\section{Introduction}
The Dirac equation provides the fundamental quantum description of spin-$1/2$ particles and has played a central role in modern physics since its formulation in 1928~\cite{Dirac1928}. Among its landmark achievements are the prediction of antiparticles~\cite{Anderson1932} and the establishment of the theoretical foundations of quantum electrodynamics~\cite{Bjorken1965, Peskin1995}. In condensed-matter physics, Dirac-like Hamiltonians also emerge as effective low-energy descriptions, giving rise to quasiparticles with relativistic or quasi-relativistic behavior.

Since the experimental isolation of graphene in 2004, genuinely two-dimensional crystals have become experimentally accessible and controllable. In monolayer graphene, low-energy quasiparticles near the Dirac points behave as massless Dirac fermions in $(2+1)$ dimensions, providing one of the clearest realizations of relativistic-like dynamics in condensed-matter systems~\cite{Novoselov2004,CastroNeto2009}. In transition-metal dichalcogenides such as MoS$_2$ and WS$_2$, a massive Dirac structure arises together with spin-valley coupling and strong spin-orbit effects~\cite{Xu2014}. In bilayer graphene, by contrast, the low-energy spectrum is approximately quadratic in momentum, and the corresponding effective description departs from Lorentz-invariant linear dispersion~\cite{McCannKoshino2013}. These examples show that low-dimensional quantum materials provide experimentally accessible settings in which both Dirac-like dynamics and controlled departures from it can be explored.

Within an effective field theory (EFT) framework, departures from Lorentz invariance may arise from intrinsic anisotropies, lattice effects, or proximity to critical regimes with nontrivial scaling behavior. A natural way to parameterize such departures is through Lifshitz scaling, under which space and time transform anisotropically as $x \to \lambda x$ and $t \to \lambda^z t$, where $z$ is the dynamical critical exponent. The relativistic case corresponds to $z = 1$, whereas $z \neq 1$ characterizes anisotropically scaled systems commonly encountered in condensed-matter contexts~\cite{Hornreich1975,Horava2009}. Lifshitz-type extensions therefore provide a natural framework for describing nonrelativistic or anisotropic dispersion relations in two-dimensional systems.

In this work, we investigate the spectral properties of a Dirac-type equation in $(2+1)$ dimensions modified by Lifshitz spatial derivatives with dynamical exponent $z = 2$. Our main goal is to determine how the Lifshitz deformation affects energy spectra under radial confinement. To this end, we combine analytical solutions for benchmark potentials with numerical calculations based on the Numerov method and a semiclassical WKB analysis. This unified approach allows us to identify characteristic scaling laws for the energy levels in different confinement regimes.

From an effective field theory perspective, such results help clarify how higher-order spatial derivatives modify spectral properties in two-dimensional Dirac systems and may be relevant for low-energy descriptions of materials with quadratic dispersion, such as bilayer graphene and related anisotropic $2D$ systems. Despite the broad literature on Lifshitz-type field theories, the spectral properties of confined Dirac-Lifshitz systems remain comparatively unexplored.

The paper is organized as follows. In Sec.~\ref{sec:model}, we introduce the Lifshitz-type Dirac equation and derive the stationary Hamiltonian. In Sec.~\ref{sec:benchmark}, we analyze the constant-potential case and radial hard-wall confinement. In Sec.~\ref{sec:osc}, we solve the harmonic-confinement problem and obtain the corresponding energy spectrum. In Sec.~\ref{sec:log}, we study logarithmic confinement numerically through the Numerov method and analytically through a semiclassical WKB treatment. Finally, in Sec.~\ref{sec:conclusions}, we summarize our main results and discuss possible extensions. In Appendix \ref{A}, we transform the logarithmic problem into a form suitable for the Numerov and WKB methods.


\section{The model}
\label{sec:model}

We consider a Dirac-type equation in $(2+1)$ dimensions modified by Lifshitz spatial derivatives, which explicitly break Lorentz invariance through higher-order spatial operators. The model is defined by
\begin{equation}
\left[i\gamma^0 \partial_0 + b\,(i\gamma^i \partial_i)^z - M \right]  \mathit{\Psi}(t, \mathbf{r}) = 0,
\end{equation}
where $\gamma^\mu$ are Dirac matrices in a $2 \times 2$ representation, $b$ is a positive deformation parameter with dimensions $[b] = [\text{mass}]^{1-z}$, $M$ is a positive mass scale, and $ \mathit{\Psi}$ is a two-component spinor field. 
Throughout this work, we adopt the representation $(\gamma^0,\gamma^1,\gamma^2) = (\sigma_z,i\sigma_x,i\sigma_y)$, where $\sigma_i$ are the Pauli matrices. We restrict attention to the case $z = 2$, for which the higher-order spatial term produces a quadratic dependence on spatial derivatives, characteristic of Lifshitz-type dispersion. To obtain the stationary Hamiltonian, we multiply the equation by $\gamma^0$ and use the Clifford algebra $\{\gamma^\mu,\gamma^\nu\} = 2g^{\mu\nu}$, where $g^{\mu\nu}$ is the Minkowski metric, with the indices $\mu$ and $\nu$ taking the values $(0,1,2)$, which yields

\begin{equation}
(i\gamma^i \partial_i)^2 = \nabla^2,
\end{equation}
where $\nabla^2 = \partial_1^2 + \partial_2^2$ is the two-dimensional Laplacian.

The equation of motion can then be written as
\begin{equation}
i\partial_0 \mathit{\Psi}(t, \mathbf{r}) = \gamma^0 (-b\nabla^2 + M) \mathit{\Psi}(t, \mathbf{r}).
\end{equation}
Introducing an external scalar potential $V(\mathbf{r})$, the stationary Hamiltonian becomes
\begin{equation}
\label{eq:Hamiltonian}
H = \gamma^0 \left(-b\nabla^2 + M \right) + V(\mathbf{r}),
\end{equation}
which defines the eigenvalue problem $H \mathit{\Psi}(t, \mathbf{r}) = E \mathit{\Psi}(t, \mathbf{r})$, where $E$ is an energy eigenvalue. 

The inclusion of external potentials in the Dirac equation is well established and respects the structure of the Poincaré group: the minimal coupling 
$p_\mu \to p_\mu - eA_\mu$ introduces a vector-type interaction \cite{Dong2003}, whereas the substitution $M\to M+S(r)$ introduces a Lorentz-scalar interaction~\cite{Lunardi2025, Hosseinpour2017}. These couplings generally produce distinct effects on the particle--antiparticle spectrum. In the present effective model, however, Lorentz invariance is explicitly broken by the $z=2$ Lifshitz spatial operator, so the usual relativistic classification does not carry over straightforwardly~\cite{Horava2009}. For the free kinetic sector, the remaining continuous symmetries are time and spatial translations together with planar rotations, $SO(2)$. After the introduction of a radial potential, spatial translations are broken, while rotational symmetry is preserved. We therefore adopt the simplest local interaction compatible with this residual symmetry, namely a multiplicative radial potential $V(\mathbf r)\mathbb{I}$ added to $\gamma^0(-b\nabla^2 + M)$, which leads to the Hamiltonian in Eq.~\eqref{eq:Hamiltonian}. 

In the absence of external potentials, the model leads to the dispersion relation
\begin{equation}
E_\pm(k) = \pm \left(M + b k^2\right).
\end{equation}

This dispersion differs from both the relativistic Dirac spectrum and the nonrelativistic Schrödinger form. In particular, the energy depends linearly on the mass scale while exhibiting a quadratic dependence on momentum, reflecting the underlying $z = 2$ scaling. The spectrum should therefore be interpreted within an effective low-energy framework in which higher-order spatial derivatives encode departures from Lorentz-invariant dynamics.

\begin{figure} [h]
    \centering
    \includegraphics[width=1\linewidth]{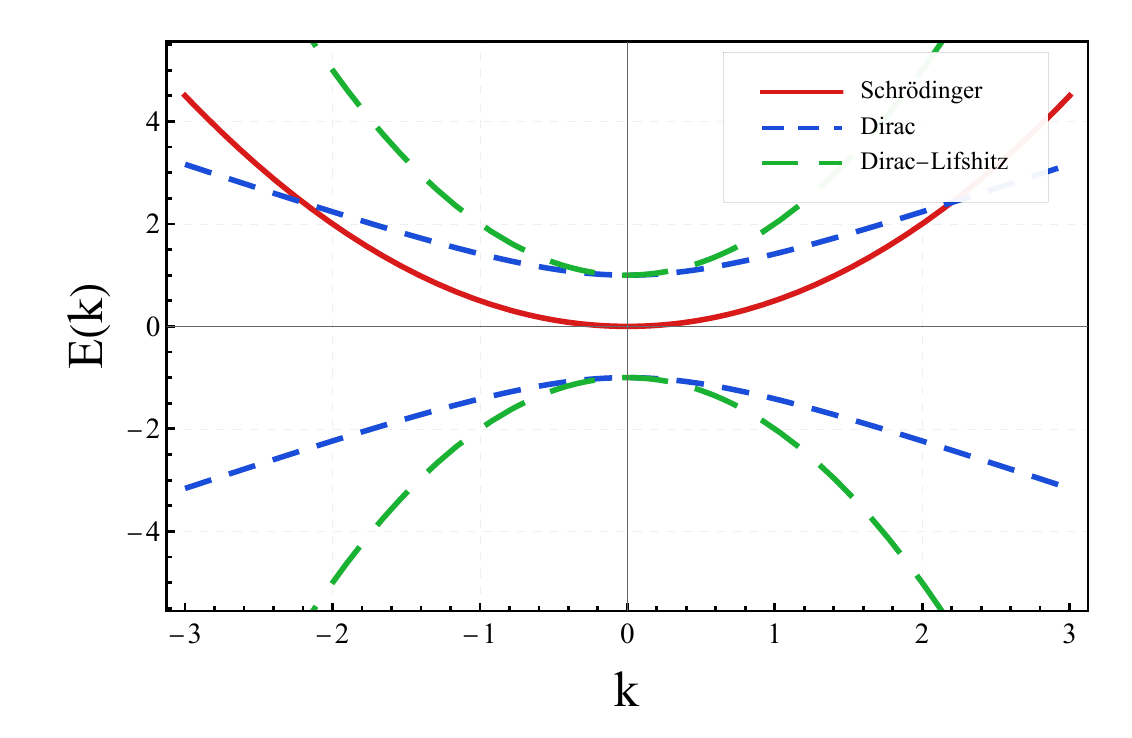}
    \caption{Qualitative representations of dispersion relations $E(k)$ for the Schrödinger, Dirac, and  Dirac-Lifshitz Hamiltonians. The parameters used were $M = 1 u\Lambda$ for all cases and $b = 1 u\Lambda^{-1}$.  The Dirac and Dirac-Lifshitz models show symmetric positive and negative branches separated by a finite gap $|\Delta| = 2M$ at $k = 0$, while the Schrödinger case shows a single quadratic branch.}
    \label{fig:Dispersion}
\end{figure}

To illustrate the qualitative features of this dispersion, we introduce an energy scale $\Lambda$. Consequently, the mass is expressed as $u \Lambda$ (in units of $\Lambda$), while the parameter $b$ is written as $u \Lambda^{-1}$ (in units of inverse $\Lambda$). In this representation, the Lifshitz-type spectrum can be compared with the standard Schrödinger dispersion $E_S(k) = k^2/2M$ and the relativistic Dirac spectrum $E_D(k) = \pm \sqrt{k^2 + M^2}$. As shown in Fig. \ref{fig:Dispersion}, the Dirac-Lifshitz model exhibits symmetric energy branches separated by a finite gap $2M$, while displaying a quadratic momentum dependence distinct from both limiting cases.

In two dimensions, such a quadratic dispersion leads to a density of states that differs qualitatively from the Dirac case. This feature suggests that the Lifshitz deformation captures effective regimes closer to systems with quadratic or anisotropic band structures, such as bilayer graphene or related materials~\cite{Katsnelson,Kotov,Lee2024}.

For time-independent Hamiltonians in Eq.~\eqref{eq:Hamiltonian}, we separate temporal and spatial variables through the ansatz
\begin{equation}
\mathit{\Psi}(t, \mathbf{r}) = e^{-iEt}\mathbf{\Psi}(\mathbf{r}),
\end{equation}
which reduces the problem to a stationary eigenvalue equation.

Since $\gamma^0$ is diagonal in the chosen representation, its eigenvalues naturally split the spectrum into positive- and negative-energy branches. Accordingly, we write the spinor as
\begin{equation}
\mathbf{\Psi}(\mathbf{r}) =
\begin{pmatrix}
\psi_1(\mathbf{r}) \\
\psi_2(\mathbf{r})
\end{pmatrix}.
\label{eq:spinor}
\end{equation}
Due to the structure of Eq.~\eqref{eq:Hamiltonian}, the functions $\psi_1$ and $\psi_2$ are decoupled for any potential term of the form $V(\mathbf r)\mathbb{I}$, where $\mathbb{I}$ denotes the identity matrix.

Considering a radial potential $V(r)$, we then apply the method of separation of variables in polar coordinates~\cite{griffiths2018introduction} and express the spinor component as follows
\begin{equation}
\psi_{a}(\mathbf{r})=R_{a}(r)\ \Phi(\phi),
\label{eq:spinorsep}
\end{equation}
where $a={1,2}$ denotes the spinor components. In polar coordinates, the Laplacian operator takes the form
\begin{equation}
\nabla^{2}=\frac{1}{r}\frac{\partial}{\partial r}
\left(r\frac{\partial}{\partial r}\right)+\frac{1}{r^{2}}\frac{\partial^{2}}{\partial\phi^{2}}.
\label{eq:laplacian-polar}
\end{equation}

Then, the angular equation becomes the following
\begin{equation}
\Phi''(\phi)+ \ell^{2}\Phi(\phi)=0,
\label{eq:angular}
\end{equation}
with integer angular momentum $\ell= \{ 0,\pm 1,\pm 2,\ldots \} $. Single-valuedness implies
\begin{equation}
\Phi_{\ell}(\phi)=A\,\e^{\dot{\imath} \ell\phi},
\label{eq:Phi}
\end{equation}
where $A$ is a normalization constant.

On the other hand, the radial equation reads
\begin{equation}
R_{a}''(r) +\frac{1}{r}R_{a}'(r)-\frac{\ell^{2}}{r^{2}}R_{a}(r) 
+\left[\frac{\mathcal{E}_{a}-M}{b}\right]R_{a}(r)=0.
\label{eq:radial2}
\end{equation}
where $\mathcal{E}_{a}$ denote the positive- and negative-energy branches, respectively, i.e., $\mathcal{E}_{1}=E-V(r)$ and $\mathcal{E}_{2}=-E + V(r)$.

Equation~\eqref{eq:radial2} has the form of a Schrödinger-like radial equation with an energy-dependent effective potential. In the present case, however, the Lifshitz term modifies the kinetic sector itself, rather than reproducing the standard relativistic Dirac structure.

In the next section, we analyze the solution of Eq.~\eqref{eq:radial2} in two cases: at constant potential and for a finite potential well. Furthermore, we compare the radial solutions with those of the Schrödinger and Dirac equations.

\section{Constant background and radial confinement}
\label{sec:benchmark}

\subsection{Free particle and constant potential}

We first consider a constant potential background, $V(r) = V > 0$. Although this case does not exhibit a bound-state spectrum, it provides a simple analytically tractable reference case. To this end, we solve the radial equation, Eq. (\ref{eq:radial2}), for each spinor component independently. The resulting general solution can be expressed in terms of linear combinations of Bessel functions  given by 
\begin{equation}\label{eq:SolBesselFunc}
R_a(r) = d_{a1} \,J_{|\ell|}(C_a r) + d_{a2} \,Y_{|\ell|}(C_a r),
\end{equation}
where $J_{|\ell|}$ and $Y_{|\ell|}$ are Bessel functions of the first and second kinds, respectively. The label $\ell$ corresponds to the angular momentum quantum number, while $d_{a1}$ and $d_{a2}$ are arbitrary constants to be determined by the boundary conditions, and $C_a$ is the effective wavenumber for the $a$ component, given by 
\begin{equation}
    C_{a}=\sqrt{\frac{\mathcal{E}_{a}-M}{b}}.
\label{eq:Cpm}
\end{equation}
Since $C_a$ is real for $\mathcal{E}_a>M$, regularity at the origin requires $d_{a2}=0$. Therefore, the physical radial solution becomes
\begin{equation}
R_a(r) = d_{a1} J_{|\ell|}(C_a\, r).
\label{eq:BesselRegular}
\end{equation}
This solution corresponds to scattering states. Furthermore, the condition $\mathcal{E}_2 > M$ characterizes the negative-energy branch, corresponding to physical energies satisfying $E < V(r)-M$.

On the other hand, for $\mathcal{E}_a<M$, the parameter $C_a$ becomes purely imaginary. In this case, Eq.~\eqref{eq:BesselRegular} can be rewritten in terms of modified Bessel functions. The resulting solutions may be either real or complex, depending on the value of $\ell$. Furthermore, they either exhibit a monotonically increasing behavior or diverge at the origin and, therefore, do not satisfy the required physical boundary conditions. Consequently, no nontrivial physically acceptable solution exists, and the only admissible solution is $R_a(r)=0$.

For this potential, the energy spectrum is separated by a forbidden gap defined by $V(r)-M < E < V(r)+M$. States with energies outside this interval correspond to free-particle-like solutions. Thus, the role of the potential is to shift the gap from the interval $[-M,M]$ to $[V(r)-M, V(r)+M]$.
To distinguish the two allowed energy sectors, we introduce the notation $E_1$ for energies satisfying $E>V(r)+M$ and $E_2$ for energies satisfying $E<V(r)-M$. Accordingly, we write
\begin{equation}
\begin{split}
\mathit{\Psi}_\ell(t,r, \phi)
= A\,\mathrm{e}^{i \ell \phi}
\Bigg[
d_{11} J_{|\ell|}\left(C_1 \ r\right) \mathrm{e}^{-i E_1 t}
\begin{pmatrix}
1\\
0
\end{pmatrix}
\\[2mm]
+\, d_{21} J_{|\ell|}(C_2 \ r) \mathrm{e}^{-i E_2 t}
\begin{pmatrix}
0\\
1
\end{pmatrix}
\Bigg].
\end{split}
\end{equation}

\begin{figure} [h]
    \centering
    \includegraphics[width=1\linewidth]{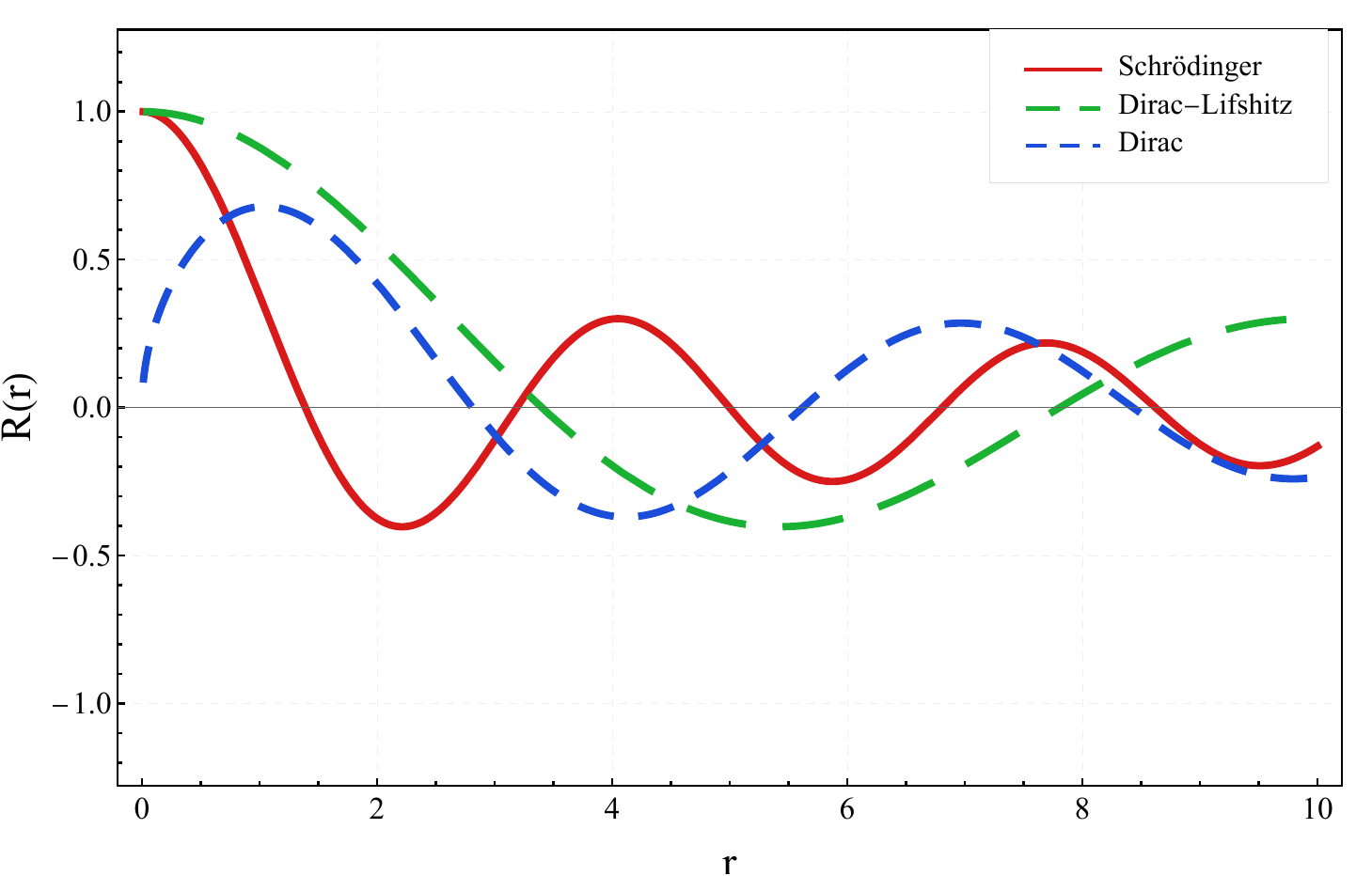}
    \caption{Comparison of the radial functions for each model with $V=0$. The chosen parameters are $\mathcal{E}_1=E_S=E_D=1.5 u\Lambda$, $M = 1u\Lambda $ and $\ell=0$ in all cases; and for the Lifshitz parameter $b = 1u\Lambda^{-1}$. These choices allow for a direct comparison of their oscillatory structure.}
    \label{fig:RadialFunction}
\end{figure}

To highlight the distinctive radial behavior of the model, we qualitatively compare Eq.~\eqref{eq:BesselRegular} for $V = 0$ with the corresponding solutions of the radial Schrödinger and Dirac equations. In particular, we consider $R_S(r) \sim J_{|\ell|}(k_S\ r)$ for the Schrödinger case and $R_D(r) \sim J_{|\ell-1/2|}(k_D \ r)$ for the Dirac case, where $k_S = \sqrt{2 M E_S}$ and $k_D = \sqrt{E_D^2 - M^2}$, as shown in Fig.~\ref{fig:RadialFunction}.

As illustrated in Fig.~\ref{fig:RadialFunction}, the Dirac-Lifshitz model displays an oscillatory radial structure whose node positions and characteristic scale are controlled by the parameter $b$. Although qualitative similarities with the Schrödinger and Dirac cases can be identified, the correspondence is not exact, especially in the presence of external potentials.

This behavior indicates that the Dirac-Lifshitz Hamiltonian defines a distinct spectral regime characterized by quadratic dispersion and a modified radial structure, rather than a simple interpolation between relativistic and nonrelativistic dynamics. 
For the free case, the choice $b=(\mathcal{E}_a-M)/(2ME_S)$ reproduces the Schrödinger curve. Furthermore, for each energy branch, a corresponding value of $b$ can be identified that yields the Schrödinger limit for $V=0$ and $\mathcal{E}_a>M$. Although qualitative similarities can be observed, the correspondence with these limiting regimes is not exact, particularly in the presence of external potentials.

This analysis highlights that the Dirac-Lifshitz Hamiltonian defines a distinct spectral regime characterized by quadratic dispersion and modified radial structure, rather than a simple interpolation between relativistic and nonrelativistic dynamics.

\subsection{Radial hard-wall confinement (``finite well'' benchmark)}

We now consider radial confinement inside a disk of radius $R_0$, implemented through hard-wall boundary conditions. This setup provides a simple benchmark for confined two-dimensional systems, such as quantum dots, in which the wave function vanishes at $r = R_0$.

It is important to note that the spinor possesses nonvanishing components inside the barrier for $\mathcal{E}_a>M$.

Imposing the boundary condition $R_a(R_0)=0$ leads to the quantization condition
\begin{equation}
J_{|\ell|}\!\left(\sqrt{\frac{\mathcal{E}_a - M}{b}}\,R_0\right)=0\,.
\end{equation}
Denoting by $j_{\ell,n}$ the $n$-th zero of $J_{|\ell|}$, the modulus of energy spectrum is given by
\begin{equation}
|E_{n,\ell}| = M + b\left(\frac{j_{\ell,n}}{R_0}\right)^2.
\label{eq:hardwall-spectrum}
\end{equation}
The energy spectrum consists of two branches, corresponding to the eigenvalues $|E_{n,\ell}|$ and $-|E_{n,\ell}|$. The corresponding eigenfunctions take the form
\begin{equation}
\begin{split}
\mathit{\Psi}_{n,\ell}(t,r, \phi)
=& A_{n,\ell}\, \mathrm{e}^{i \ell \phi} J_{|\ell|}\!\left(\frac{j_{\ell,n}}{R_0} r\right) \times \\
& \times \Bigg[
\mathrm{e}^{-i |E_{n,\ell}| t}
\begin{pmatrix}
1\\
0
\end{pmatrix}
+\, \mathrm{e}^{i |E_{n,\ell}| t}
\begin{pmatrix}
0\\
1
\end{pmatrix}
\Bigg]  ,
\end{split}
\end{equation}
where $A_{n,\ell}$ is a normalization constant.

A key feature of this spectrum is the scaling
\begin{equation}
|E_{n,\ell}| - M \propto \frac{b}{R_0^2},
\end{equation}
which directly reflects the anisotropic character of the Dirac-Lifshitz model. This quadratic dependence on the inverse confinement radius contrasts with the linear scaling typically found in massless Dirac systems. The result shows that the Lifshitz parameter $b$ sets the confinement-induced energy scale and therefore provides a clear spectral signature of anisotropic scaling.

The eigenfunctions form a complete orthogonal set within the disk and can serve as a basis for the analysis of more general potentials or perturbations.

This benchmark also provides a useful point of comparison for smooth confining potentials, such as the harmonic oscillator considered in Sec.~\ref{sec:osc}, where the quantization condition follows from the analytic structure of the wave function rather than from an imposed hard boundary.

\section{Harmonic oscillator}
\label{sec:osc}
We now consider quadratic confinement described by the harmonic potential
\begin{equation}\label{eq:OH_Potential}
V(r) = \frac{1}{2}\,\omega^2 r^2.
\end{equation}

This potential provides a useful reference model for smooth confinement in two-dimensional systems and still allows for an analytical treatment within the present framework. 
 
In this work, the potential is introduced through Eq.~\eqref{eq:Hamiltonian}. This choice implies that only one component of the spinor in Eq.~\eqref{eq:spinor} corresponds to a physical solution for the potential given by Eq.~\eqref{eq:OH_Potential}, namely $\psi_{1}(\mathbf{r})$. This can be understood by analyzing the radial equation, Eq.~\eqref{eq:radial2}. As discussed below Eq.~\eqref{eq:radial2}, each value of the index $a$ is associated with a different energy branch of the potential. In particular, $\mathcal{E}_1 \propto -r^2$, whereas $\mathcal{E}_2 \propto +r^2$.

To clarify this point, let us analyze the two radial solutions separately. For $a=1$, the coefficient of $r^2$ is negative, yielding an asymptotically attractive harmonic potential that forces the radial function to decay as a Gaussian, guaranteeing square‑integrability. For $a=2$, however, the coefficient of $r^2$ is positive, turning the potential into an inverted oscillator \cite{Barton1986}.

In general, the inverted oscillator is associated with non-normalizable states, a continuous spectrum, and an imaginary frequency \cite{Barton1986}. Some of its applications include quantum chaos \cite{QuChenLiu2022}, the study of statistical fluctuations in fission dynamics \cite{Hofmann1997}, and the construction of tachyonic backgrounds in string theory \cite{Sera2005}. Under certain boundary conditions, this model may exhibit discrete energy levels whose energy eigenvalues decay over time \cite{Yuce2006}. However, in the present work, we focus on bound-state stationary solutions with real energy eigenvalues. Consequently, the radial solution $R_1(r)$ is the only physically acceptable nontrivial solution.

To solve the radial equation for $a =1$ in the harmonic potential case, let is define $\kappa^2 = (E - M)/b$ and $\Omega^2 = \omega^2/(2b)$. We then obtain
\begin{equation}
    R_1''(r)+\frac{1}{r}R_1'(r) + \left[\kappa^2 - \Omega^2 r^2 - \frac{\ell^2}{r^2}  \right]R_1(r) = 0.
\label{eq:ohq1}
\end{equation}
Introducing the dimensionless variable $\rho = \sqrt{\Omega}\, r$, the equation becomes
\begin{equation}
R_1''(\rho)+ \frac{1}{\rho}R_1'(\rho) + \left(\frac{\kappa^2}{\Omega} - \rho^2 - \frac{\ell^2}{\rho^2}\right) R_1(\rho) = 0.
\label{differencialpho}
\end{equation}

The asymptotic behavior of Eq.~\eqref{differencialpho} determines the structure of the physical solutions. For $\rho \to 0$, the centrifugal term dominates and yields $R_1(\rho) \sim \rho^{|\ell|}$, whereas for $\rho \to \infty$ the harmonic term dominates and the solution decays as $R_1(\rho) \sim \mathrm{e}^{-\rho^2/2}$.

These asymptotic behaviors motivate the ansatz
\begin{equation}
R_1(\rho) = \rho^{|\ell|} \mathrm{e}^{-\rho^2/2} F\left(\rho\right).
\end{equation}
Substituting this ansatz into Eq.~\eqref{differencialpho}, one finds that $F(\rho)$ satisfies a confluent hypergeometric equation. Upon introducing the variable $x = \rho^2$, the equation becomes

\begin{multline}
    x F''(x) + (|\ell| + 1 - x) F'(x) + \\ \frac{1}{4}\left(\frac{\kappa^2}{\Omega} - 2|\ell| - 2\right) F(x) = 0,
\end{multline}
whose regular and normalizable solutions are the associated Laguerre polynomials $\mathrm{L}_n^{|\ell|}(x)$, with $n =\{0,1,2,\ldots\}$.

The normalized radial eigenfunctions are therefore given by
\begin{equation}
\mathbf{\Psi}_{n,\ell}(r,\phi) = N_{n,\ell}\ \mathrm{e}^{-\frac{1}{2}\Omega r^2+\ii \ell\phi}\, r^{|\ell|}  \ \mathrm{L}_n^{|\ell|}\left(\Omega r^2\right)
\begin{pmatrix}
1 \\
0
\end{pmatrix},
\label{eq:osc-radial}
\end{equation}
where $N_{n,\ell}$ is a normalization constant. The requirement of polynomial solutions imposes the quantization condition
\begin{equation}
\frac{\kappa^2}{\Omega} = 4n + 2|\ell| + 2,
\end{equation}
which determines the discrete spectrum. Since Eq.~\eqref{eq:ohq1} requires $\kappa^2 > 0$, it follows that $E > M$. Consequently, the spectrum contains only positive-energy states. This structure reflects the standard two-dimensional harmonic oscillator \cite{villalba1994exact,moshinsky1989dirac}.

Using the definitions of $\kappa$ and $\Omega$, the energy spectrum is therefore given by
\begin{equation}
E_{n,\ell}= M+\sqrt{2b}\,\omega\left(2n+|\ell|+1\right).
\label{eq:osc-spectrum}
\end{equation}

This result shows that the Lifshitz parameter $b$ controls the level spacing through the effective scale $\sqrt{2b}$, reflecting the underlying $z = 2$ anisotropic scaling. The harmonic case therefore provides a useful analytically solvable example of smooth confinement within the Dirac-Lifshitz framework.

\section{Logarithmic confinement: numerical spectrum and semiclassical scaling}
\label{sec:log}

We now consider the logarithmic confining potential
\begin{equation}\label{eq:PotentialLog}
V(r) = \alpha \ln(r/r_0),
\end{equation}
where $\alpha$ is a positive parameter that sets the interaction strength and $r_0$ defines a reference scale. Potentials of this type arise in effective low-dimensional descriptions and provide a useful setting for investigating semiclassical quantization.

For the logarithmic potential defined in Eq.~\eqref{eq:PotentialLog}, Eq.~\eqref{eq:radial2} does not belong to any of the known classes of exactly solvable differential equations. Nevertheless, it can be treated by several analytical and numerical methods, such as numerical integration and asymptotic approximations. Among these, the WKB approximation~\cite{Froman1965} provides a natural framework for analyzing its asymptotic behavior.

As in the previous case, for the $a=2$ branch, Eq.~\eqref{eq:radial2} with the logarithmic potential admits no bound-state solutions. In the asymptotic limit $r\to\infty$, Eq.~\eqref{eq:radial2} reduces to
\begin{equation}
    R_2''(r)+\frac{\alpha}{b}\ln\left(\frac{r}{r_0}\right)R_2(r) \approx0\,.
\end{equation}
The corresponding asymptotic solutions are purely oscillatory, with amplitudes decaying only as $(\ln r)^{-1/4}$. Consequently, $R_2(r)$ is not square-integrable and therefore cannot represent a bound state.

On the other hand, the $a=1$ branch exhibits a discrete energy spectrum. To determine the corresponding solutions $R_1(r)$, we employ the Numerov method~\cite{hairer1993nonstiff}. The resulting spectrum is then compared with the semiclassical WKB approximation~\cite{Froman1965}.
For the numerical treatment, it is convenient to rewrite the radial equation. By introducing the transformation $R_1(r) = u(r)/\sqrt{r}$, the first-derivative term is removed, and the radial equation is cast into a Schrödinger-like form; we obtain the following equation,
\begin{equation}
u''(r) + Q(r)\,u (r) = 0,
\label{QLN}
\end{equation}
where
\begin{equation}
Q(r) = \frac{1}{b}\left[E - M - \alpha \ln(r/r_0)\right] - \frac{\ell^2 - 1/4}{r^2}.
\label{eq:Q}
\end{equation}
Details of this reduction are given in the Appendix \ref{A}.

\subsection{Numerov method}

The reduced radial equation is solved numerically with the Numerov method, which is particularly suitable for second-order differential equations of the form $u''(r) + Q(r)u(r) = 0$ with sufficiently smooth $Q(r)$.

The discretized recurrence relation is given by
\begin{equation}
u_{i+1} = \frac{\left(2 - \frac{5\tau^2}{6}Q_i\right)u_i - \left(1 + \frac{\tau^2}{12}Q_{i-1}\right)u_{i-1}}{1 + \frac{\tau^2}{12}Q_{i+1}},
\label{eq:numerov}
\end{equation}
where $\tau$ denotes the radial step size and $Q_{i}=Q(r_{i};E,\ell)$.

The eigenvalues are determined through a shooting procedure by selecting solutions that remain finite near the origin and decay at large distances~\cite{blatt1967practical}. The radial function is normalized according to
\begin{equation}
\int_0^\infty |R_1(r)|^2 r\,dr = 1.
\end{equation}

In the numerical implementation, the radial domain is chosen large enough to ensure convergence of the bound-state eigenvalues. The resulting discrete spectrum for $\ell = 0$ is shown in Fig.~\ref{fig:ComparaçãoWKBxNumerov}.



\subsection{WKB quantization and Lifshitz scaling law}

To obtain analytical insight into the spectrum, we employ a semiclassical WKB approximation. The corresponding quantization condition reads~\cite{desai2010quantum}

\begin{equation}
\int_{r_-}^{r_+} \sqrt{Q(r)}\,dr = \pi\left(n + \frac{1}{2}\right),
\label{I36}
\end{equation}
where $n=\{0,1,2,\ldots\}$ and $r_\pm$ are the classical turning points. In WKB, the region $Q(r)>0$ corresponds to oscillatory solutions (classically allowed region), while $Q(r)<0$ corresponds to exponential decay. 

In the large-$n$ regime, the centrifugal term may be neglected to leading order~\cite{Froman1965,Karnakov2012}, yielding
\begin{equation}
I(E) = \int_{r_-}^{r_+} \sqrt{\frac{E - M - \alpha \ln(r/r_0)}{b}} \, dr.
\label{I37}
\end{equation}

This approximation becomes asymptotically accurate for highly excited states, for which the turning points occur at sufficiently large $r$ and the logarithmic potential dominates over the centrifugal contribution.


Introducing the change of variable
\begin{equation}
s=E-M-\alpha\ln (r/r_0),
\label{eq:change}
\end{equation}
we obtain
\begin{equation}
r=r_0\exp\left(\frac{E-M-s}{\alpha}\right),
\label{eq:rfroms}
\end{equation}
and consequently
\begin{equation}
\dd r=-\frac{r_0}{\alpha}\exp\left(\frac{E-M-s}{\alpha}\right)\dd s.
\label{eq:dr}
\end{equation}

Substituting Eqs.~\eqref{eq:change}–\eqref{eq:dr} into Eq.~\eqref{I37} leads to
\begin{equation}
I(E)=\frac{r_0}{\alpha \sqrt{b}}\exp\left(\frac{E-M}{\alpha}\right)
\int_{0}^{E-M} s^{1/2}\, \mathrm{e}^{-s/\alpha}\dd s.
\label{eq:I2}
\end{equation}
The remaining integral has the structure of an incomplete gamma function. Introducing $L = E - M$, it can be written as
\begin{equation}
    J_{\text{exact}} = \int_0^{L} \!\! s^{1/2}\ \mathrm{e}^{-s/\alpha} \dd s = \alpha^{3/2} \ \gamma\left(\frac{3}{2},\frac{L}{\alpha} \right),    
\end{equation}
where $\gamma(q,x)$ denotes the lower incomplete gamma function. In the semiclassical regime, corresponding to large radial quantum numbers $n \gg 1$, the energy satisfies $L = E - M \gg \alpha$. Under this condition, the exponential factor $\exp(-s/\alpha)$ decays rapidly, and the integral is dominated by values $s \sim \alpha$. Consequently, extending the upper integration limit to infinity introduces only a small error. The integral can therefore be approximated by
\begin{equation} 
    J_{\text{approx}} = \int_0^{\infty} \!\! s^{1/2}\ \mathrm{e}^{-s/\alpha} \dd s = \alpha^{3/2} \ \Gamma(3/2).
\end{equation}

The difference between the exact and approximate integrals corresponds to the tail contribution
\begin{equation}
    \Delta J = J_{\text{approx}} - J_{\text{exact}} = \int_L^{\infty} \!\! s^{1/2} \ \mathrm{e}^{-s/\alpha} \dd s.
\end{equation}

This term can be expressed in terms of the upper incomplete gamma function \cite{BenderOrszag1999},
\begin{equation}
    \Delta J =  \alpha^{3/2} \ \Gamma \left( \frac{3}{2},\frac{L}{\alpha}\right).
\end{equation}

For large arguments, $x = L/\alpha \gg 1$, the upper incomplete gamma function admits the asymptotic expansion

\begin{equation}
    \Gamma(q,x) \sim x^{q-1}\, \mathrm{e}^{-x} \left[1 + \mathcal{O}(1/x) \right].
\end{equation}

Applying this result with $q = 3/2$, the relative error of the approximation becomes
\begin{equation} \frac{\Delta J}{J_{\text{approx}}} \sim \frac{2}{\sqrt{\pi}} \left( \frac{L}{\alpha} \right)^{1/2} \exp\left( -\frac{L}{\alpha} \right), 
\end{equation}
which is exponentially small. Therefore, extending the upper integration limit to infinity is an excellent approximation in the semiclassical regime, consistently with the Laplace-method estimate~\cite{BenderOrszag1999}. Performing this extension in Eq.~\eqref{eq:I2}, we obtain the approximate expression
\begin{equation}
I(E) \approx \frac{r_0}{2} \exp\left(\frac{E - M}{\alpha}\right) \sqrt{\frac{\alpha \pi}{b}}.
\label{eq:Iapprox}
\end{equation}

Returning to Eq.~\eqref{I36},
\begin{equation}
\frac{r_0}{2}\exp\left(\frac{E-M}{\alpha}\right)\sqrt{\frac{\alpha\pi}{b}}
=\pi\left(n+\frac{1}{2}\right).
\label{eq:quant}
\end{equation}

Solving for $E$ yields the semiclassical scaling law

\begin{equation}
E_{n}\approx M + \alpha \ln \! \left[\frac{2}{r_0}\sqrt{\frac{\pi b}{\alpha}}
\left(n+\frac{1}{2}\right)\right].
\label{eq:WKBEn}
\end{equation}

\begin{figure} [h]
    \centering
    \includegraphics[width=1\linewidth]{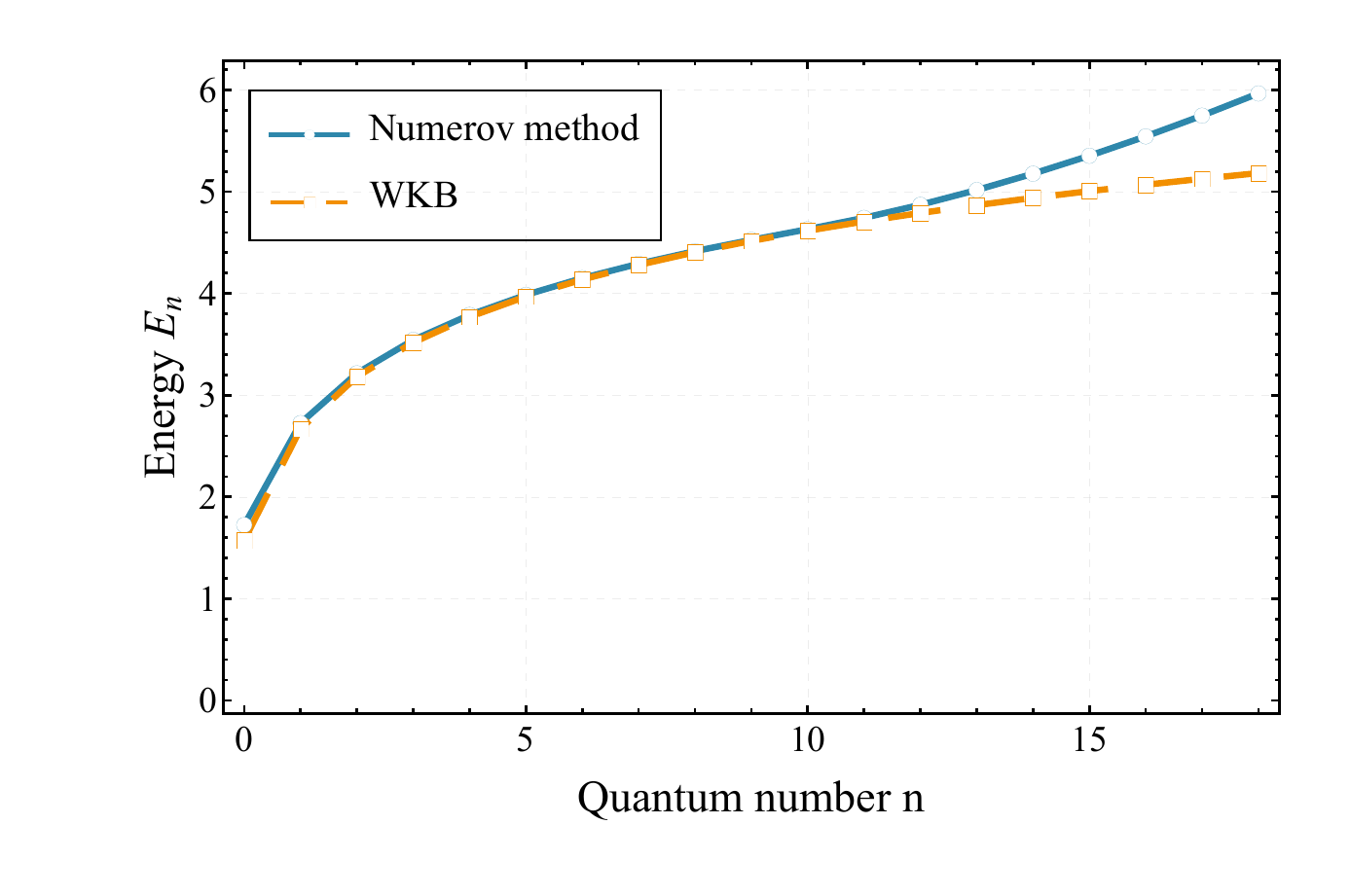}
    \caption{Comparison between the numerical spectrum obtained with the Numerov method and the semiclassical prediction of Eq.~\eqref{eq:WKBEn} for $\ell=0$, $b=1u\Lambda^{-1}$, $\alpha = 1u\Lambda$, $r_0 = 1 u\Lambda^{-1}$ and $M = 1 u\Lambda$. The radial cutoff was chosen sufficiently large to ensure the proper decay of the wave functions for highly excited states.}
    \label{fig:ComparaçãoWKBxNumerov}
\end{figure}

This result predicts a logarithmic growth of the energy levels, $E_n - M \sim \alpha \ln \sqrt{b}$, in agreement with the numerical solutions obtained via the Numerov method, as shown in Fig.~\ref{fig:ComparaçãoWKBxNumerov}.

Figure~\ref{fig:ComparaçãoWKBxNumerov} compares the numerical spectrum obtained with the Numerov method with the semiclassical prediction in Eq.~\eqref{eq:WKBEn}. For sufficiently large radial cutoffs, the two approaches are in very good agreement over the range of quantum numbers considered. In both cases, the spectrum exhibits the characteristic logarithmic growth associated with the confining nature of the potential.

\begin{figure} [h]
    \centering
    \includegraphics[width=1.0\linewidth]{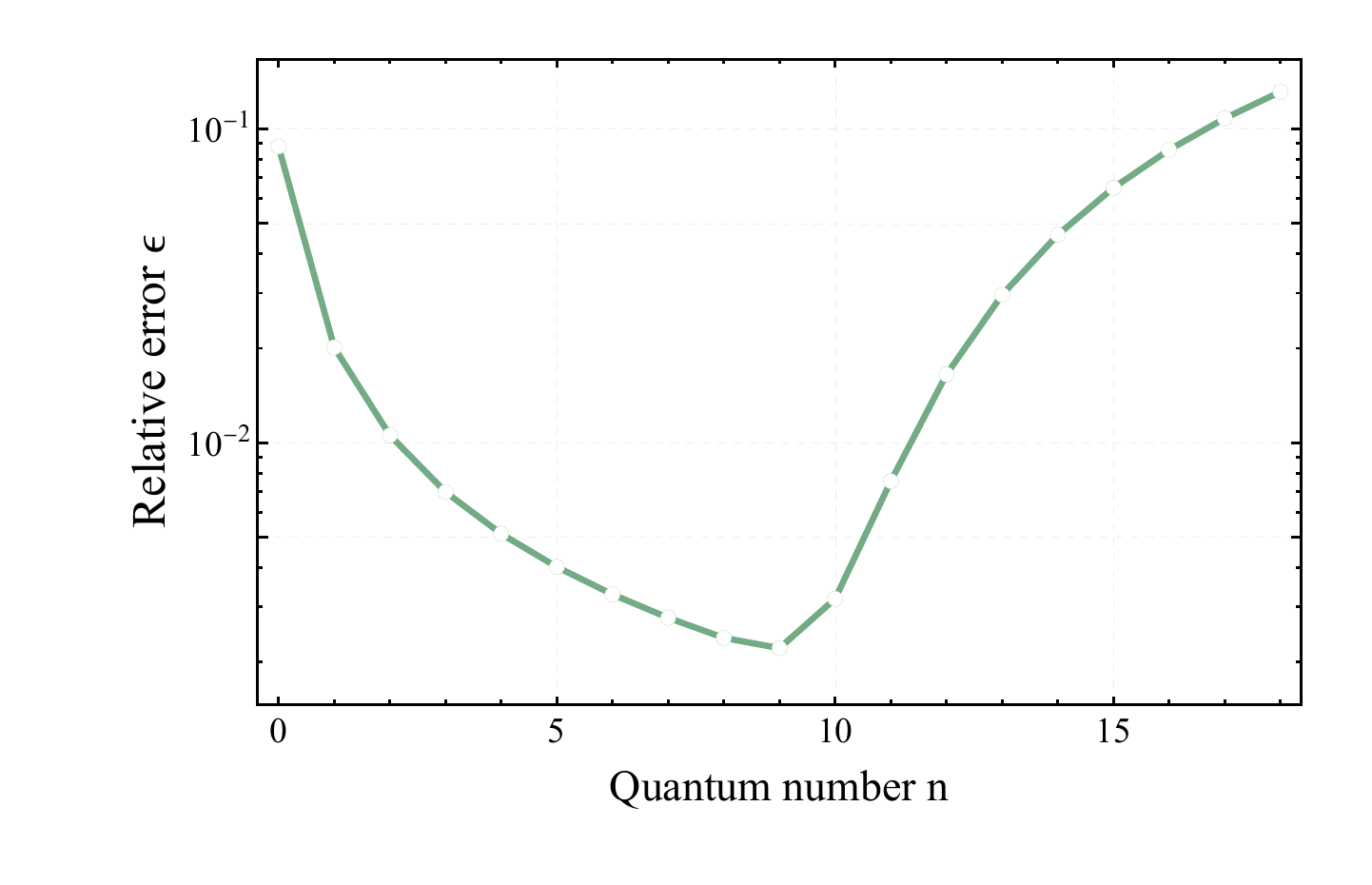}
    \caption{Relative error between the eigenvalues obtained with the Numerov method and the semiclassical estimate given by Eq.~\eqref{eq:WKBEn}. The deviation is largest for the lowest state $n = 0$ and for large-$n$, where semiclassical approximations are not expected to be reliable. As the quantum number increases the agreement improves, reaching a minimum error around $n = 9$. For higher states the error slowly increases again, which can be attributed to the growing spatial extent of the excited states and the finite radial cutoff used in the numerical integration.}
    \label{fig:RelativeError}
\end{figure}

To quantify this agreement, we evaluate the relative error $\epsilon = (E_{\mathrm{Numerov}} - E_{\mathrm{WKB}})/E_{\mathrm{Numerov}}$ as a function of the radial quantum number $n$. The results, shown in Fig.~\ref{fig:RelativeError}, identify the regime in which the WKB approximation provides an accurate description of the spectrum, as well as the range in which noticeable deviations persist.

Having verified that the WKB approximation accurately reproduces the numerical spectrum for $\ell = 0$, we now extend the analysis to states with nonzero angular momentum. In this case, the centrifugal term modifies the effective potential, as shown in Eq.~\eqref{eq:Q}, and is expected to affect the semiclassical quantization condition. Guided by the structure of radial WKB methods, we introduce an effective shift in the quantum number $n \rightarrow n + c_\ell$, where $c_\ell$ is determined by fitting the WKB expression to the numerical spectrum for $\ell \neq 0$. This procedure effectively incorporates the contribution of the centrifugal term, which is not fully captured by the leading-order logarithmic approximation.
We find that the semiclassical spectrum can be significantly improved by incorporating an effective shift in the radial quantum number, leading to the expression

\begin{equation}
E_{n,\ell}\approx M+\alpha\ln\!\left[\frac{2}{r_0}\sqrt{\frac{\pi b}{\alpha}}
\left(n+\frac{|\ell|}{2}+\frac{3}{4}\right)\right].
\label{eq:WKBEnl}
\end{equation}

This form provides an excellent agreement with the numerical results obtained via the Numerov method, with only small deviations for the lowest the highest angular momentum states.

\section{Conclusions and outlook}
\label{sec:conclusions}

We have investigated the spectral properties of a Dirac-type equation in $(2+1)$ dimensions modified by Lifshitz spatial derivatives with dynamical exponent $z = 2$, with emphasis on energy spectra under radial confinement. The higher-order spatial term produces a quadratic dispersion relation and, consequently, systematic departures from Lorentz-invariant Dirac dynamics.

Analytical solutions were obtained for constant backgrounds, hard-wall confinement, and harmonic potentials, whereas logarithmic confinement was studied through a combination of Numerov-based numerical calculations and a semiclassical WKB analysis. Across these regimes, the energy spectrum exhibits characteristic scaling laws, including $(E - M) \propto b/R_0^2$ for hard-wall confinement, $E - M \propto \sqrt{2b}\,\omega$ for harmonic trapping, and $E - M \sim \alpha \ln \sqrt{b}$ for logarithmic confinement in the semiclassical regime. The agreement between the numerical and semiclassical results supports the consistency of these spectral behaviors.

From an effective field theory perspective, these results provide a consistent characterization of how Lifshitz-type modifications affect energy spectra in two-dimensional Dirac systems. Such behavior may be relevant for effective descriptions of systems with parabolic or anisotropic low-energy dispersion.

Future work may extend the present analysis by incorporating external fields, boundary effects, and interactions, which may help clarify the role of Lifshitz-type operators in more realistic two-dimensional settings.

\begin{acknowledgments}
L.\ K.\ Q.\ and N.\ B.\ were partially supported by the Coordena\c{c}\~ao de Aperfei\c{c}oamento de Pessoal de N\'ivel Superior (CAPES). V.\ S.\ A.\ was partially supported by the Conselho Nacional de Desenvolvimento Cient\'ifico e Tecnol\'ogico (CNPq-Brazil) through Process No.\ 408735/2023-6 (CNPq/MCTI). L. F.\ F.\ A. acknowledges financial support from the Vicerrector\'ia de Investigaci\'on y Postgrado de la Universidad de La Frontera through Project No.\ BPVRIP-092024. The authors also thank Luis B. Castro (UFMA-Brazil) for carefully reading the manuscript and for pointing out some sign inconsistencies. The authors dedicate this work to the memory of Francisco Peña, whose scientific insight, collaboration, and friendship were invaluable throughout this project.

\end{acknowledgments}

\appendix

\section{Details of the radial reduction}\label{A}


Starting from Eq.~\eqref{eq:Hamiltonian} and using polar coordinates, one arrives at Eq.~\eqref{eq:radial2} after separating variables. After applying the derivatives and renaming the radial functions, we arrive at Eq.~\eqref{eq:radial2}. That way, we can make the transformation $R_1(r)=u(r)/\sqrt{r}$. This is the standard radial reduction used to transform the problem into a form suitable for the Numerov and WKB methods. Applying this transformation, we have for the first derivative
\begin{equation}
    R_1'(r) =  -\frac{1}{2} r^{-3/2}u(r) + r^{-1/2}u'(r)
\end{equation}
and differentiating again,
\begin{equation}
    R_1''(r) = \frac{3}{4}r^{-5/2}u(r) - r^{-3/2}u'(r) + r^{-1/2}u''(r).
\end{equation}
Therefore,
\begin{equation}
    R_1''(r) + \frac{1}{r}R_1'(r) = r^{-1/2}u''(r) + \frac{1}{4}r^{-5/2}u(r),
\end{equation}
where the first-derivative terms cancel.

Therefore, Eq.~\eqref{eq:radial2}, expressed in terms of the function $u(r)$, takes the form
\begin{multline}
    \left[ \frac{1}{4}r^{-5/2}u(r) + r^{-1/2}u''(r) \right] - \ell^2r^{-5/2}u(r) \\
    + \frac{1}{b}\left[E - M - \alpha\ln{(r/r_0)} \right]r^{-1/2}u(r) = 0,
\end{multline} 
factoring out $r^{-1/2}$, the reduced equation for $u(r)$ is Eq.~\eqref{QLN}. This transformation introduces the effective centrifugal term $\ell^2 - 1/4$ in the Eq.~\eqref{eq:Q}, which is purely geometric and arises from the two-dimensional radial measure. This structure is analogous to the appearance of $\ell(\ell+1)$ in three-dimensional radial problems~\cite{berry1973semiclassical}.



\end{document}